\documentclass{elsart3p}

\usepackage{epsfig}

\usepackage{amssymb}

\begin{document}

\begin{frontmatter}

\title{Anisotropic flow in Pb+Pb collisions at LHC from 
the quark gluon string model with parton rearrangement}

\author[label1]{J. Bleibel}
\author[label2]{, G. Burau}
\author[label1]{, C. Fuchs}

\address[label1]{Institut f{\"u}r Theoretische Physik, 
Eberhard Karls Universit{\"a}t, Auf der Morgenstelle 14, 
D-72076 T\"ubingen, Germany}
\address[label2]{Institut f{\"u}r Theoretische Physik, 
J. W. Goethe-Universit{\"a}t, 
Max-von-Laue-Stra{\ss}e 1, D-60438 Frankfurt am Main, Germany}

\begin{abstract}
We present predictions for the pseudorapidity dependence of the azimuthal 
anisotropy parameters $v_1$ and $v_2$ of baryons and inclusive charged 
hadrons in Pb+Pb collisions at a LHC energy of 
$\sqrt{s_{NN}} = 5.5~{\rm TeV}$ applying a microscopic transport model, 
namely the quark gluon string model (QGSM) which has been recently 
extended for parton rearrangement and fusion processes. Pb+Pb collisions 
with impact parameters $b = 2.3~{\rm fm}$ and $b = 8~{\rm fm}$ have been 
simulated in order to investigate additionally the difference between 
central and semiperipheral configurations. 
In contrast to $v_1^{\rm ch}(\eta)$ at RHIC, the directed flow of charged 
hadrons shows a small normal flow alignment. The elliptic flow 
$v_2^{\rm ch}(\eta)$ turns out to be rather similar in shape for RHIC and LHC 
conditions, the magnitude however increases about $10-20 \%$ at the LHC, 
leading to the conclusion that the hydrodynamical limit will be reached.
\end{abstract}

\end{frontmatter}


\section{Introduction}
\label{intro}

Ultra-relativistic heavy ion collisions have been performed within various
experiments at the Relativistic Heavy Ion Collider (RHIC) in Brookhaven.
Since 2000 gold on gold collisions at center of mass energies up to 
$\sqrt{s_{NN}}=200$ GeV have been investigated. After many years of operation
strong experimental evidence has been accumulated, that at these energies 
indeed a new state of matter is created, which is qualitatively different from 
a hadron gas (see \cite{QM2005} and references therein). This new state 
is believed to consist of deconfined partons, as predicted by calculations 
within Quantum Chromodynamics (QCD) on the lattice \cite{Karsch:2006sm,Aoki:2006br}. It does not behave like a weakly interacting gas of partons 
and rather exhibits features of a strongly coupled system, a strongly coupled 
Quark Gluon Plasma (sQGP). 
The strong elliptic flow signal measured at RHIC \cite{Ackermann:2000tr,Park:2001gm,Manly:2002uq,Back:2004mh,Adler:2003kt} is one of the key observables 
justifying such a scenario. A strongly interacting system would
imply large pressure gradients and short equilibration 
times \cite{Heinz:2001xi,Shuryak:2003xe}, both being necessary 
conditions for the dynamics leading to the development of strong 
elliptic flow. The scaling behavior of the elliptic flow of the final hadrons 
with the number of constituents (see e.g. Refs. 
\cite{Abelev:2007qg,Bass:2006ib}) is a second hint towards the partonic 
nature of the created medium. Assuming, that the elliptic flow is to the most 
extent already created in the partonic phase of the collision, the observed 
scaling can be naturally explained as the result flow being transfered from 
the constituent partons to the final hadrons via parton recombination or 
coalescence mechanisms \cite{Dover:1991zn,Hwa:2002tu,Greco:2003xt,Molnar:2003ff,Fries:2003kq}.

The newly built Large Hadron Collider (LHC) at CERN in Geneva is intended 
to start operation in 2008. Among the various experiments, the dedicated 
Heavy Ion Program of the ALICE Collaboration \cite{Carminati:2004fp} intends 
to investigate Pb+Pb Collisions at center of mass energies up to 
$\sqrt{s_{NN}}=5500$ GeV. This increase in energy of more than an order 
of magnitude as compared to RHIC offers the opportunity to study the 
properties of the strongly coupled quark gluon plasma more closely, 
since the energy density and lifetime of the partonic system will increase
\cite{Eskola:2005ue}. For the anisotropic flow, especially the elliptic flow, 
the question whether the hydrodynamical limit will be finally reached or not 
will be of particular interest. At RHIC energies, it has recently been 
concluded that this limit is reached only to a level of $\approx 70-80\%$ 
\cite{Hirano:2007gc,Drescher:2007cd}. This lack of perfection of the 
``perfect liquid'', especially at higher rapidities can be seen for example 
in the pseudorapidity dependence of elliptic flow \cite{Hirano:2001eu}, 
which cannot be described in terms of ideal hydrodynamic. 
So far the description of this data at RHIC has been achieved only after the 
inclusion of a dissipative hadronic cascade in a Hydro-Cascade hybrid model 
\cite{Hirano:2005xf} and with a partonic rearrangement ansatz within the 
microscopic quark gluon string model \cite{Bleibel:2006xx}.

LHC predictions for elliptic flow from hydrodynamical calculations as in 
Refs. \cite{Hirano:2007gc,Snellings:2006qw} as well as from scaling 
arguments \cite{Borghini:2007ub} show a further increase of elliptic flow, 
which is in line with further approach to the hydrodynamical limit, whereas 
a parton transport approach \cite{Molnar:2007an} predicts a significant
smaller anisotropy parameter $v_2$. In the present work we apply the 
aforementioned quark gluon string model with parton rearrangement for lead on 
lead collisions at top LHC energy and two different impact parameters, 
namely $b=2.3$ fm corresponding to the mean impact parameter of the $5\%$ most 
central collisions and $b=8$ fm as a representative impact parameter for 
semiperipheral collisions. Thus, we are able to present predictions for 
the pseudorapidity dependence of directed and elliptic flow for inclusive 
charged hadrons and inclusive baryons for both centralities.

\section{QGSM with parton rearrangement}
\label{mcqgsm}
As basis for the present study serves the Monte-Carlo version of the 
quark gluon string model (QGSM) \cite{QGSM0,QGSM1c} which has been recently 
extended in order to allow for parton exchange (rearrangement) and fusion 
processes \cite{Bleibel:2006xx}. The standard version of the QGSM, i.e. the 
model without partonic rearrangements, incorporates already partonic and 
hadronic degrees of freedom and is based on Gribov-Regge theory accomplished by
a string phenomenology of particle production in inelastic hadron-hadron 
collisions. Thus, strings in the QGSM can be produced as a result of 
color exchange (Pomeron exchange) and, like in diffractive scattering, 
due to momentum transfer. Hard gluon-gluon scattering and semi-hard 
processes with quark and gluon interactions have been also incorporated 
in the model \cite{hard}. The cascade procedure of multiple secondary 
interactions of produced hadrons was implemented in order to describe 
hadron-nucleus and nucleus-nucleus collisions. QGSM and other string-cascade 
models have been successfully applied to describe directed and elliptic flow at
SPS energies \cite{LPX99,Petersen:2006vm,DF_prc00,DF_prc01}. Also at RHIC, 
the bulk properties of elliptic flow have been fairly well reproduced within 
the standard version of the QGSM \cite{Burau:2004ev,Zabrodin:2005pd}. 
In addition it has been shown that energy densities well above critical values 
predicted by lattice QCD are achieved with the QGSM, and corresponding energy 
density profiles at proper time $\tau = 1~{\rm fm/c}$ compare well with 
hydrodynamical assumptions for initial distributions \cite{Bleibel:2005gx}.

As mentioned above, the QGSM describes particle production by the 
excitation and decay of open strings with different partons, 
namely (anti)quarks or (anti)diquarks, on their ends. 
Therefore it has provided a framework for the inclusion of 
partonic rearrangement processes which can occur in the very dense 
stages of a heavy ion reaction where the ``hadrons'' overlap and consequently 
are not really bound states anymore, but rather strongly correlated 
quark-antiquark or (anti)quark-(anti)diquark states. 
Please note, that in contrast to Ref. \cite{Bleibel:2006xx}, we call the 
extension of our model \emph{parton rearrangement} here, in order to clearly 
distinguish this ansatz from the well established ``parton recombination'' and
``parton coalescence'' models \cite{Hwa:2002tu,Greco:2003xt,Molnar:2003ff,Fries:2003kq}. 
So, the idea is basically the following: Above a critical local 
(energy/particle) density, ``hadrons'' satisfying corresponding constrains are 
decomposed into their constituent partons which then are allowed to rearrange 
themselves into new ``hadronic correlations''. Additionally, a quark-antiquark 
pair of the same flavor may annihilate during the rearrangement process with a 
given probability, implementing effectively a $3\to 2$ reaction. The 
probability for these annihilation processes was fixed in 
Ref. \cite{Bleibel:2006xx} for RHIC energies and has now been extrapolated to 
LHC, assuming a weak dependence on the center of mass energy of the collision:
\begin{equation}
P_a(\sqrt{s_{NN}})=0.04\,\sqrt{\frac{s_{NN}}{s_{NN}^{RHIC}}}^\lambda,\quad
\lambda=0.288
\end{equation}
By means of that, this ansatz takes the increased likelihood for $3\to 2$ 
reactions due to the increased particle density into account. It was 
motivated by the so called pocket formula \cite{Armesto:2004ud}
derived within a saturation model. However, any centrality dependence of the 
annihilation probability has been neglected. 

Partonic rearrangement or annihilation processes might very 
frequently happen as long as the local density of the medium is high enough. 
Accordingly, these rearrangement processes become more and more unimportant 
when the system increasingly thins out. In this spirit the model effectively 
emulates a medium of very strongly coupled partons, i.e. quark-antiquark and 
(anti)quark-(anti)diquark states, during the early times of an 
ultra-relativistic heavy ion collision. We want to note that this model 
does not create a system of ``free partons''. Insofar the QGSM upgraded by 
the locally density dependent parton rearrangement mechanism quasi 
models - within its limitations - the possible dynamics of a sQGP from a 
microscopical point of view \cite{Bleibel:2006xx}.

\section{Anisotropic flow at LHC}
\label{flow}

Collective flow phenomena are among the main signals, which can help to 
reveal the formation of the sQGP in the experiment. Flow is directly linked 
to the equation of state of the excited matter produced in ultra-relativistic 
heavy ion collisions. One can subdivide the transverse collective flow 
into isotropic and anisotropic flow. The two most important types of 
anisotropic flow are characterized by the first and the second harmonic 
coefficients of the Fourier decomposition of the invariant azimuthal 
particle distribution in momentum space \cite{VoZh96,PoVo98}:
\begin{equation}
E \frac{d^3N}{d^3p} = \frac{1}{\pi} \frac{d^2N}{dp_T^2dy}
\left[
1 + 2 \sum_{n=1}^{\infty} v_n(p_T,y) \cos(n\phi)
\right]
\label{distribution}
\end{equation}
Here, $p_T=(p_x^2+p_y^2)^{1/2}$ is the transverse momentum, $y$ the
rapidity and $\phi$ the azimuthal angle of a particle between its 
momentum and the reaction plane.

For the following study of anisotropic flow, we simulated $Pb+Pb$ 
collisions at top LHC energy. For the central collisions, the number of
charged hadrons at midrapidity reaches up to $dN_{ch}/d\eta\,|_{\eta=0}= 
3813$, whereas in the semiperipheral case the simulation only yields 
$dN_{ch}/d\eta\,|_{\eta=0}= 953$. 

\begin{figure}[ht]
  \centering
  \epsfig{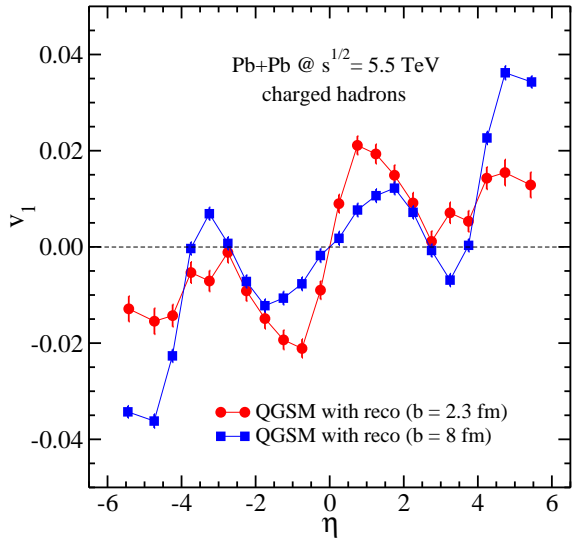}
  \caption{\label{v1etaHchar}Directed flow $v_1$ of charged hadrons 
  as a function of pseudorapidity $\eta$ for Pb+Pb collisions at 
  $\sqrt{s_{\rm NN}} = 5.5~{\rm TeV}$ with impact parameters $b = 2.3~{\rm fm}$ 
  and $b = 8~{\rm fm}$, respectively. Statistical errors are indicated 
  by bars.}
\end{figure}

The first harmonic coefficient $v_1$ in Eq. (\ref{distribution}) is called 
directed flow given by $v_1 = \langle \cos(\phi) \rangle = 
\langle p_x/p_T \rangle$. 
Its pseudorapidity dependence $v_1(\eta)$ for inclusive charged hadrons 
extracted from the QGSM simulations is shown in Fig. \ref{v1etaHchar}. 
This first anisotropic component seems to show a small normal flow alignment, 
i.e. a positive slope $dv_1/d\eta$, at mid-pseudorapidity in contrast to the 
findings at the highest RHIC energy of $\sqrt{s_{\rm NN}} = 200~{\rm GeV}$ 
\cite{Back:2005pc,Adams:2003zg} where the directed flow is essentially flat 
and close to zero in the pseudorapidity region $|\eta| \le 2$. However, for the
semiperipheral collisions also $v_1$ at LHC seems to be very small, i.e. 
less than $1.5\%$, in a broad pseudorapidity range. The structure of the 
directed flow remains the same for the most central collisions, however, the 
flow coefficient reaches higher values at even somewhat smaller rapidities. 
The slope is therefore even steeper in central collisions. 
This rather unexpected increase of $v_1$ compared to RHIC can be understood
if one considers the different viscosities in the region with ($|\eta|<3$) and 
at higher pseudorapidities. At midrapidity, the viscosity is 
very low due to the small mean free path of the particles undergoing 
rearrangement processes. At higher rapidities, the partonic rearrangement is
suppressed, therefore the viscosity is higher and due to shadowing
the antiflow components become visible. A medium with very low viscosity will
rather preserve its primordial flow -- which is normal flow i.e. 
$dv_1/d\eta>0$, since the interactions which cause the rise of the antiflow 
component (shadowing) are strongly suppressed. This also explains the larger 
direct flow for central collisions.

\begin{figure}[b]
  \centering
  \epsfig{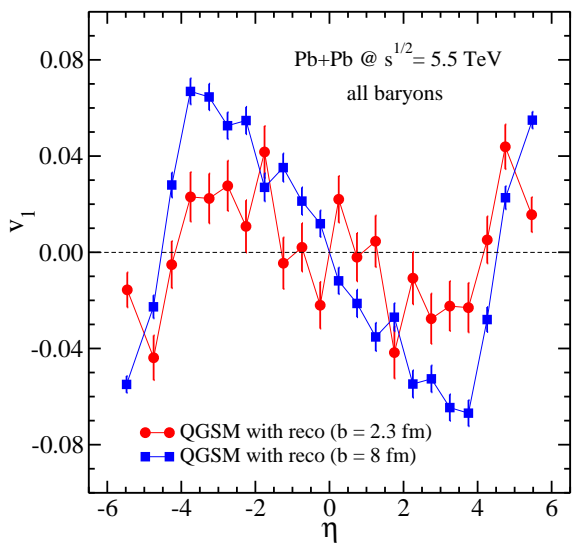}
  \caption{\label{v1etaNuc}Pseudorapidity distribution of $v_1$ for 
  baryons only. The simulated collisions are the same as in 
  Fig. \ref{v1etaHchar}.}
\end{figure}
Since the spectrum of final hadrons is dominated by pions, we separately 
show only the directed flow of all baryons in Fig. \ref{v1etaNuc}. 
Interestingly, the baryonic $v_1$ shows a negative slope $dv_1/d\eta$, 
conventionally called antiflow, for at least the semiperipheral collisions. 
At higher values of $|\eta|$, the directed flow of all baryons is 
rather large. The situation is not so clear in more central reactions. 
Here, the magnitude of the directed flow is compatible with zero at least
in the pseudorapidity region $|\eta|<1.5$, but then peaking at around 
$|\eta|\approx 2-3$. However, the statistical errors are still too large to 
draw a definite conclusion. 
In both cases the flow of baryons shows the same behavior as in our
previous study at RHIC energies \cite{Burau:2004ev}. It is less affected
by the rearrangement processes which play only a minor role for baryons.

\begin{figure}[htb]
  \centering
  \epsfig{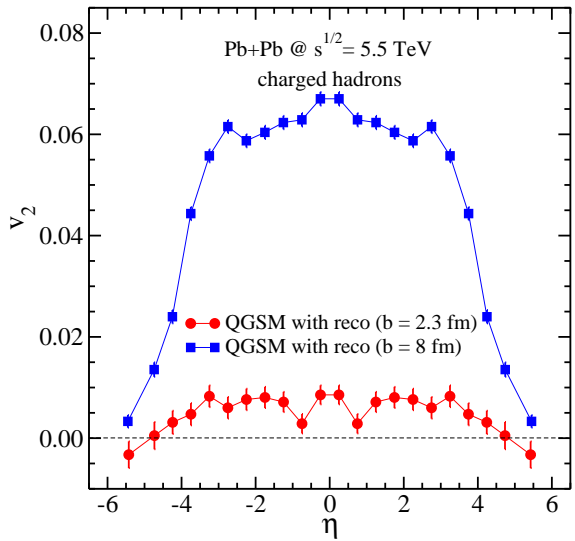}
  \caption{\label{v2etaHchar}Elliptic flow $v_2$ vs. pseudorapidity $\eta$ 
  of inclusive charged hadrons from Pb+Pb simulations with the QGSM extended 
  by parton rearrangement processes. The error bars denote statistical 
  uncertainties.}
\end{figure}

Next we investigate the pseudorapidity dependence of the elliptic flow 
$v_2(\eta)$, i.e. the second harmonic coefficient of the Fourier decomposed 
invariant azimuthal distribution of produced particles given by 
Eq. (\ref{distribution}). This anisotropic flow component is determined by 
$v_2 = \langle \cos(2\phi) \rangle = 
\langle (p_x/p_T)^2 - (p_y/p_T)^2 \rangle$. 
The QGSM simulation results for Pb+Pb collisions at 
$\sqrt{s_{\rm NN}} = 5.5~{\rm TeV}$ with impact parameters $b = 2.3~{\rm fm}$ 
and $b = 8~{\rm fm}$ are depicted in Fig. \ref{v2etaHchar}. 
The crucial result of an analogous study for Au+Au collisions at RHIC with 
the QGSM \cite{Bleibel:2006xx}, which has been extended by a locally density 
dependent partonic rearrangement mechanism in order to model effectively the 
dynamics of a very strongly coupled quark plasma at high particle densities, 
was that the shape of the anisotropy parameter $v_2(\eta)$ of final charged 
hadrons is intimately related to the collision dynamics. It turned out within 
this microscopic investigation that fast equilibration due to parton 
rearrangement and fusion processes, which occur in the very dense medium 
created in ultra-relativistic heavy ion reactions during the early stages, 
are necessary in order to obtain $v_2(\eta)$ profiles which are peaked at 
midrapidity as seen in the RHIC data. Hence it might be no surprise to find 
a very similar qualitative behavior for LHC conditions with a much higher 
collision energy.

Here, the extended QGSM predicts for semiperipheral collisions also a strong 
in-plane alignment of $v_2$ with a peak at $|\eta| \approx 0$ and a steady 
decrease for larger values of $|\eta|$, but the total distribution is of 
course broader compared to $v_2(\eta)$ at RHIC. Furthermore, the maximum value 
of the elliptic flow around midrapidity of $v_2(\eta=0)=6.7\%$ is significantly
higher than at the highest RHIC energy: The impact parameter $b=8$ fm 
corresponds to a centrality of the collision of $\approx 25\%$, or a mean 
number of participants of $N_{part}=180$. Such a maximum value of 
($p_t$- integrated) $v_2$ is not observed at RHIC even in more semiperipheral 
collisions, i.e. $25-50\%$, $N_{part}\approx 111$ \cite{Back:2004mh}. A 
simulation with the extended QGSM at maximum RHIC energy, analogous to Ref.
\cite{Bleibel:2006xx} but with the same number of participants as above, 
yielded a value of $v_2(\eta=0)=5.9\pm 0.2\%$. The predicted maximum value at 
LHC is therefore about $10-20\%$ greater than at RHIC.  This result is in line 
with the assumption, that the hydrodynamical limit was not reached at RHIC. In 
the hydrodynamical regime the elliptic flow would scale with the eccentricity 
$\epsilon$ of the overlap of the colliding nuclei, or $v_2/\epsilon=const.$. 
However, such a scaling cannot be found (see e.g. 
Refs. \cite{Hirano:2007gc,Drescher:2007cd}), the hydrodynamical limit was 
therefore reached only up to $70-80\%$. The predicted increase of elliptic flow
at LHC can in this context be interpreted as a much closer approach to the 
hydro-limit. It has been shown that microscopic transport calculations can 
indeed, at least in 2D, approach this limit \cite{Gombeaud:2007ub}. In our 
model, the convergence towards the hydrodynamical limit can be nicely explained
by the effect of the viscosity of the medium on $v_2$: due to the many partonic
rearrangements the mean free path in the medium of produced particles
is reduced. This also lowers the viscosity, and by having more and more 
rearrangement processes due to the increased energy- and particle density at 
LHC, the mean free path and thus the viscosity may actually be minimized. 
This would lead to a maximum value for the elliptic 
flow \cite{Bhalerao:2005mm}. 

A second argument for approaching the hydrodynamical limit can be 
drawn from the ratio $v_4/v_2^2$. It has been argued in Refs. 
\cite{Bhalerao:2005mm,Borghini:2005kd} that this ratio probes the 
degree of equilibration of the produced matter, leading to $v_4/v_2^2 \to 1/2$
in the hydrodynamical limit. Preliminary results for this observable, 
averaged for $0.15 < p_t < 2.0$ GeV and $|\eta| < 4$, yield for charged
hadrons a value of $v_4/v_2^2\approx 0.76 \pm 0.24$. Despite the 
large errors, the decrease from the measured $v_4/v_2^2 = 1.2$ at RHIC 
\cite{Adams:2003zg} and $v_4/v_2^2\approx 1.29 \pm 0.27$ as result of a 
similar analysis at top  RHIC energy with the same number of participants and 
based on the data of Ref. \cite{Bleibel:2006xx}, supports a 
further approach of the hydrodynamical limit and is in line with transport 
calculations by Ko et al. \cite{Ko:2007}.

The centrality dependence of $v_2(\eta)$ is qualitatively similar
for RHIC and LHC conditions: The elliptic flow is large 
for peripheral reactions and rather small for central ones. The result for the
central collisions yields a rather flat distribution of $v_2$ over a broad
centrality range with values around $v_2\approx 1\%$.
Whether the distribution is peaked or not cannot be decided due to the
limited statistics.

\begin{figure}[ht]
  \centering
  \epsfig{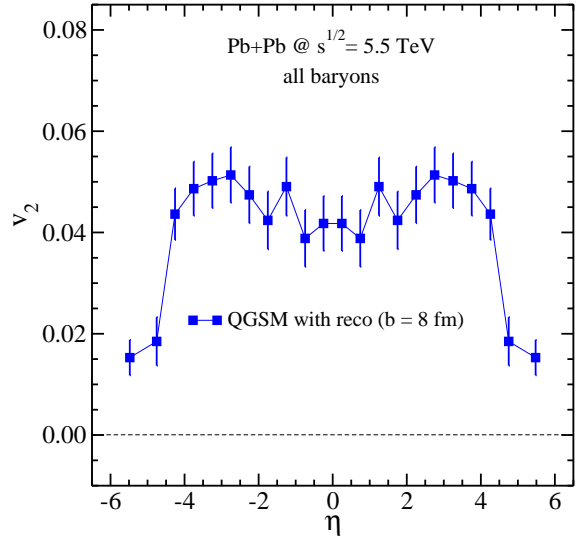}
  \caption{\label{v2etaNuc}The same like in Fig. \ref{v2etaHchar}, but for 
  inclusive charged nucleons only.}
\end{figure}

For the sake of completeness, Fig. \ref{v2etaNuc} shows the pseudorapidity 
distributions of $v_2$ for all baryons only, which have been extracted from 
the aforementioned Pb+Pb collisions at $\sqrt{s_{\rm NN}} = 5.5~{\rm TeV}$ 
with an impact parameter of $b = 8~{\rm fm}$.
The distribution for all baryons shows a different structure. In contrast to 
the single peaked pseudorapidity distribution of elliptic flow of charged 
hadrons here two peaks at pseudorapidities $|\eta|\approx 3$ are predicted. 
The values of $v_2$ at higher rapidities are quite similar to the result with 
all charged hadrons. At midrapidity, the flow is significantly smaller.

\section{Summary and conclusions}

For this survey, we have analyzed simulated Pb+Pb collisions at a center 
of mass energy of $\sqrt{s_{NN}} = 5.5~{\rm TeV}$ with impact parameters 
$b = 2.3~{\rm fm}$ and $b = 8~{\rm fm}$ applying a microscopic string-cascade 
transport model, namely the quark gluon string model (QGSM), which has been 
recently extended for locally density dependent parton rearrangement and 
fusion processes in order to emulate effectively a medium of very strongly 
correlated partons, i.e.quark-antiquark and (anti)quark-(anti)diquark 
states, and its dynamics. Predictions for the pseudorapidity dependence 
of the azimuthal anisotropy parameters $v_1(\eta)$ and $v_2(\eta)$ of nucleons 
and inclusive charged hadrons for central and semiperipheral collision 
configurations have been presented. The directed flow of charged hadrons 
shows a small normal flow alignment at midrapidity for both the central 
and semiperipheral reactions in contrast to the findings at RHIC, but it 
is rather small at all, i.e. less than $2\%$. 
The elliptic flow of charged final hadrons turns out to be large for 
semiperipheral Pb+Pb collisions with a maximum value of about 6.7\% and small 
for the central ones. The second anisotropy parameter seems to be rather 
similar in shape for RHIC and LHC conditions, the maximum value at midrapidity
however has been predicted to increase by about $10-20\%$. Therefore it has 
been concluded that the hydrodynamical limit is quasi reached.

\section*{Acknowledgments}

The authors thank L.V. Bravina and E.E. Zabrodin for fruitful discussions. 
This work has been supported by the Bundesministerium f{\"u}r Bildung und 
Forschung (BMBF) under contract 06T\"U202.



\end{document}